\documentclass[prb,twocolumn]{revtex4}

\usepackage{amssymb,amsmath,bm,epsfig,graphicx,subfigure}

\pdfoutput=1

\begin{document}

\title{Commensurate structures in twisted transition metal dichalcogenide heterobilayers}
\author{Madeleine Phillips and C. Stephen Hellberg}
    \email{steve.hellberg@nrl.navy.mil}
    \affiliation{Naval Research Laboratory, Washington, D.C. 20375 \\}
\date{\today}

\begin{abstract}
A major theoretical challenge of studying twisted transition metal dichalcogenide (TMD) bilayers is that the unit cell of such structures is very large and therefore difficult to address using first-principles methods.  However, twisted TMD bilayers form moir\'e patterns, which consist of regions of commensurate stacking, either smoothly interpolated into one another or separated by sharp domain walls. In this paper, we study twisted TMD bilayers by studying the properties of the constituent commensurate structures.  Using density functional theory (DFT), we compute band structures for commensurately-stacked MoS$_2$/WS$_2$ and MoSe$_2$/WSe$_2$ bilayers in both $0^\circ$ and $60^\circ$ orientations, and we highlight variations in band structures across different commensurate geometries.  These band structure variations arise from diverse factors such as metal atom asymmetry between layers (Mo vs. W), differences in interlayer hybridization, and Brillouin zone alignment.  We comment on the consequences of such band structure differences for optical experiments and on the effects of strain on moir\'e pattern electronic structure.     
\end{abstract}

\maketitle

In the advent of two-dimensional materials research, the most recent epoch has focused on stacks of 2D materials, known as van der Waals heterostructures \cite{NiuHetReview, GeimVDWHet, LiuHetReview}.  These heterostructures are characterized by strong bonds in the plane of each 2D layer and weak van der Waals bonds between layers.  The weak interlayer coupling makes it possible to relax constraints related to lattice-matching and interlayer alignment and create stable heterostructures out of disparate materials in a wide variety of stacking orientations.  Because heterostructure properties change when interchanging materials and altering interlayer geometry, this freedom opens the door for the creation of ``designer materials," whose properties can be tuned by careful choices of constituent layers and interlayer orientation.  

One specific degree of freedom that has attracted much attention recently is the relative twist angle between layers.  Famously, superconductivity emerges in bilayer graphene twisted at the ``magic angle" of $1.05^\circ$ \cite{CaoMagicAngle}, and topologically protected propagating states are pinned at the domain walls in bilayer graphene with small twist angles on the order of $0.25^\circ$ \cite{ZhangPNAS2013, LeRoyBLG2018}.  In transition metal dichalcogenide (TMD) bilayers, the moir\'e pattern in off-commensurate samples, i.e. those twisted a small angle away from a commensurate stacking orientation, has been invoked to explain the photoluminescence signal of interlayer excitons \cite{TranMoire2019, JinMoire2019, SeylerMoire2019}.  

A challenge of interpreting experiments carried out on small twist angle bilayers is that such systems have a very large repeat unit cell, making them prohibitively expensive to model using first-principles methods.  This is in contrast to commensurately stacked bilayers, which can be described by a unit cell containing just twice as many atoms as the monolayer unit cell. One way to address this challenge becomes apparent when we note that the moir\'e patterns of small twist angle heterostructures contain distinct regions of commensurate stacking\cite{YaoInterlayer2017, TongTopMosaic, XuHetReview2018}.  Even better, Carr et al. have recently shown that homobilayers with sufficiently small twist angles should reconstruct into configurations dominated by commensurately stacked regions, which are separated only by narrow domain walls and nodes \cite{Carr2018}.  This reconstruction has been observed extensively in bilayer graphene \cite{McEuenPNAS2013, BasovBLG2018, KimBLG2019}, and we expect the same type of reconstruction to occur in both homo- and heterobilayer TMD structures. However, whether the transitions between commensurately stacked regions are smooth (as in the rigid moir\'e structure) or sharp (as in the reconstructed moir\'e), we can learn much about the electronic properties of twisted heterobilayers by studying the constituent commensurately stacked structures.   

\begin{figure}
  \includegraphics[angle=0,width=0.9\columnwidth]{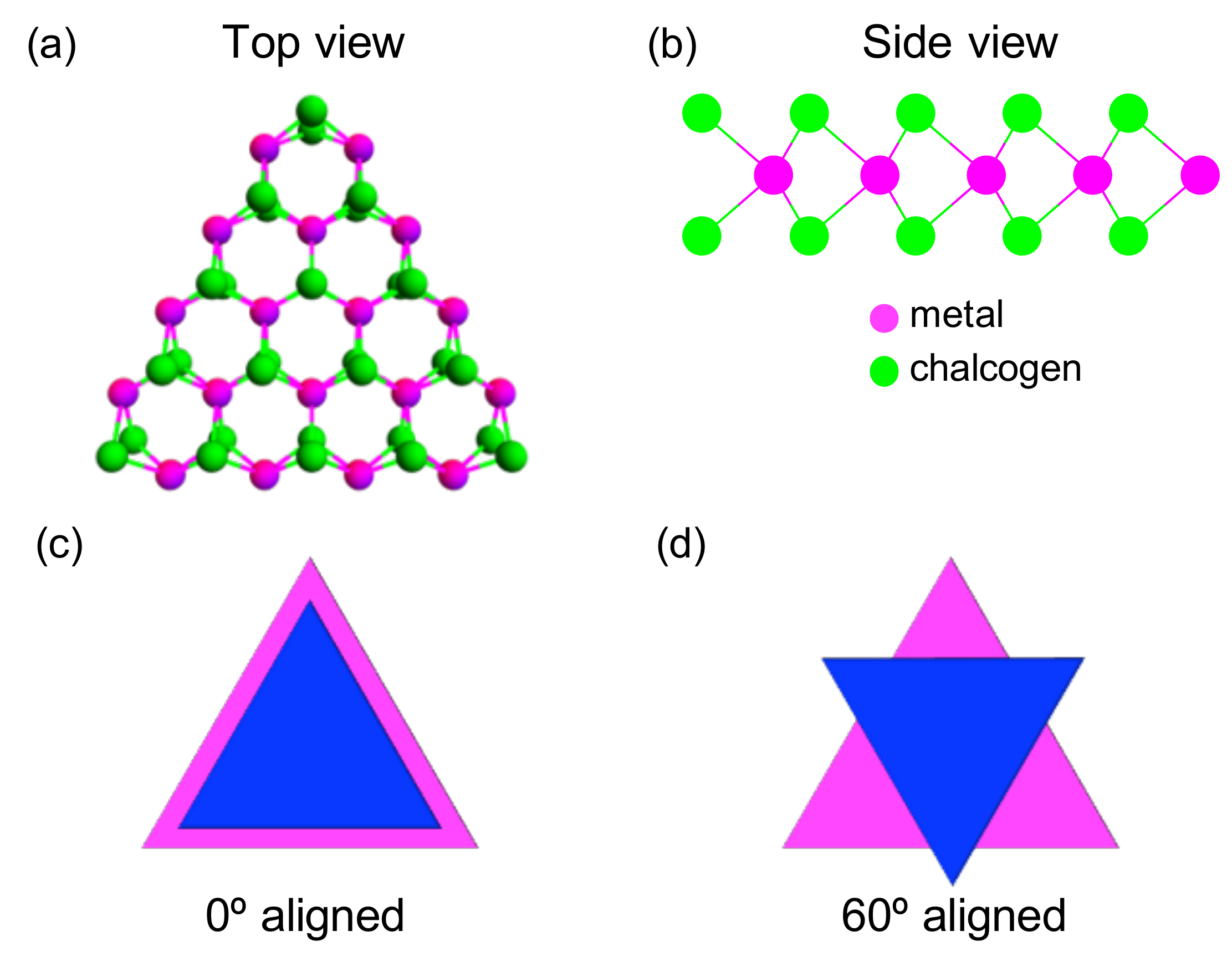}
  \caption{\label{Geom}(a) Schematic of a TMD monolayer in the 1H phase\cite{Notation}. (b) Side view of a 1H TMD monolayer.  Schematics of (c) $0^{\circ}$ and (d) $60^{\circ}$-aligned TMD bilayers. }
\end{figure}

For monolayers with hexagonal lattices, such as 1H-TMD monolayers \cite{Notation}, the commensurately stacked bilayer structures have layers oriented with a relative $0^\circ$ or $60^\circ$ twist (Figure \ref{Geom}).  The geometries accessed by translating one layer of a $0^\circ$ or $60^\circ$-oriented bilayer with respect to the other layer are also commensurate structures.  Any of these stackings may appear in the corresponding off-commensurate twisted structures (e.g. $\theta=1^\circ$ or $\theta=59^\circ$ structures), and thus the study of the whole collection of $0^\circ$ and $60^\circ$ commensurate stackings is relevant for the study of small angle twisted bilayers \cite{YaoInterlayer2017}.  In this paper, we study the configuration spaces of geometries associated with the $0^\circ$ and $60^\circ$ stacked TMD bilayers MoX$_2$/WX$_2$, where X=S or Se.  Using first principles methods, we compute the electronic structures of the high-symmetry commensurate stackings.  We focus on the high-symmetry geometries because these constitute the ground state and high energy structures in the energy landscapes of the $0^\circ$ and $60^\circ$ stacking spaces, and they are of special relevance in reconstructed moir\'e patterns \cite{Rosenberger2019, Carr2018}.  By examining the electronic structures of different commensurate geometries and comparing bilayer to monolayer band structures, we explore how band features are related to system symmetry, interlayer distance, and the difference in metal atoms between layers.  Understanding the properties of commensurate geometries individually leaves us better equipped to predict the properties of twisted bilayers.

\section{Methods}
First-principles calculations were carried out in a density functional theory (DFT) framework using the projector augmented wave (PAW) approach\cite{Blochl1994, Kresse1999} and the generalized gradient approximation (GGA)\cite{Perdew1996} as implemented in VASP\cite{VASP}.  For the MoS$_2$/WS$_2$ bilayers, we use metal potentials with six valence electrons: $4d^55s^1$ for Mo and $5d^56s^1$ for W.  For the S potential, the $n=3$ electrons are included as valence electrons ($3s^23p^4$). For the MoSe$_2$/WSe$_2$ bilayers, the metal potentials we use have 14 valence electrons each: $4s^24p^64d^6$ for Mo and $5s^25p^65d^6$ for W, while the Se potential includes the $n=4$ electrons as valence ($4s^24p^4$).  In all cases we use at least an 8 x 8 x 1 $\Gamma$-centered k-point mesh, a plane-wave energy cutoff of 450 eV, and an out of plane lattice constant of $30 \mathring{A}$, yielding a vacuum size of about $20 \mathring{A}$.  We include the van der Waals interaction using the DFT-D3 method of Grimme\cite{Grimme2010}. 

\section{Results}

\begin{figure}
  \includegraphics[angle=0,width=0.9\columnwidth]{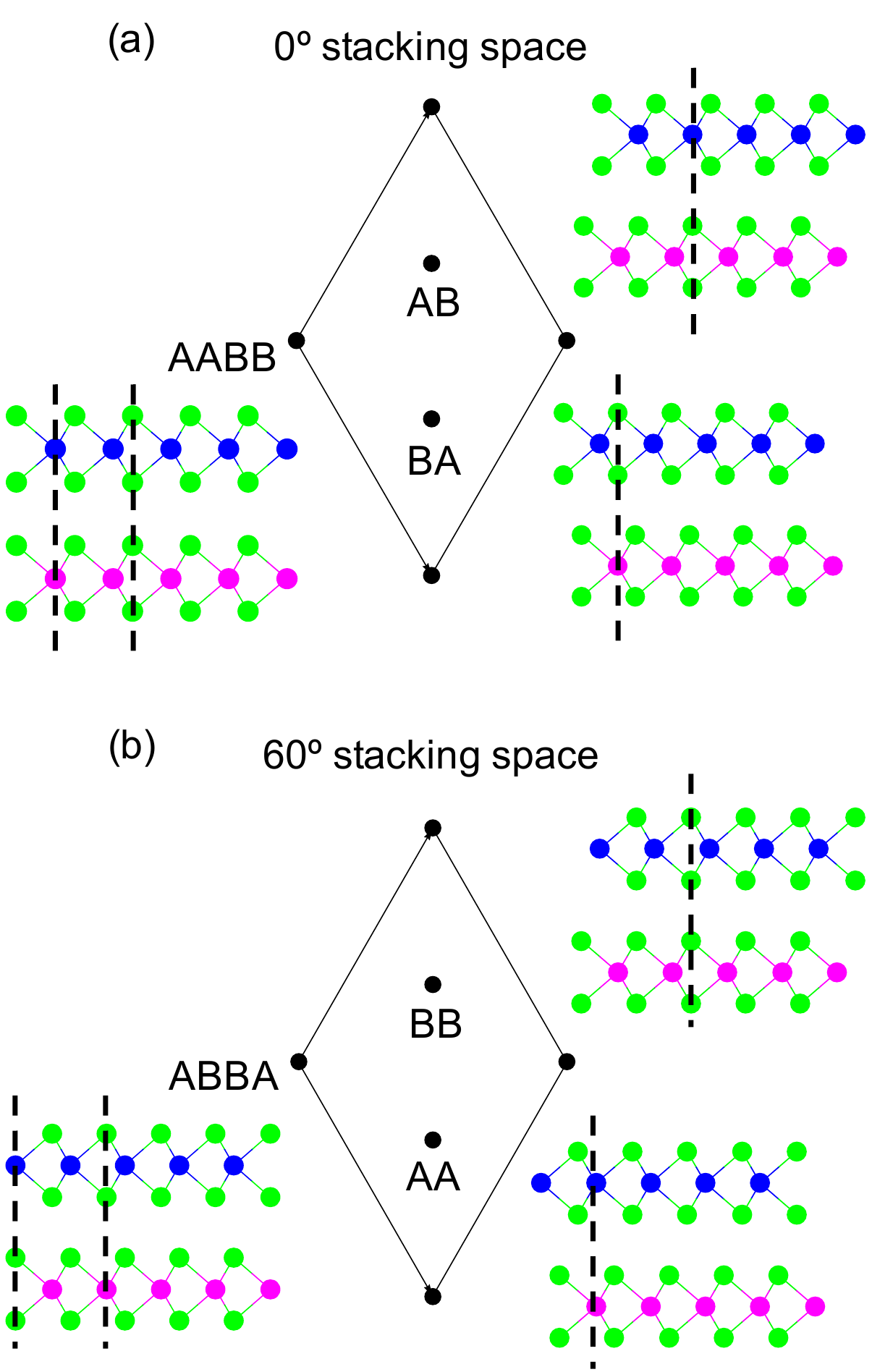}
  \caption{\label{StackGeom} Stacking spaces of (a) $0^{\circ}$- and (b) $60^{\circ}$-aligned TMD bilayers, with high-symmetry stackings labelled and depicted in side view.  High-symmetry stackings are specified by a two or four letter label, where A refers to metal sites, and B refers to chalcogen sites. The first and third letter in a label refer to sites in the top layer, and the 2nd and 4th letter in a label refer to sites in the bottom layer.  E.g. The label AABB refers to a stacking geometry where top layer metals are aligned over bottom layer metals, and top layer chalcogens are aligned over bottom layer chalcogens (A site over A site and B site over B site).}
\end{figure}

We carried out first-principles calculations for four distinct systems: MoS$_2$/WS$_2$ and MoSe$_2$/WSe$_2$ bilayers in the $0^\circ$ and $60^\circ$ stacking orientations. The geometry of each system is further specified by $\mathbf{r}(\alpha, \beta)$, which describes the lateral shift of the top layer relative to the bottom layer while keeping the angular orientation fixed. We study translations of the top layer described by the following expression: 
\begin{equation}
\label{shift}
\mathbf{r}(\alpha, \beta) = \alpha\mathbf{r}_1 + \beta\mathbf{r}_2,
\end{equation} 
where $\mathbf{r}_1 = a(\frac{\sqrt{3}}{2}, -\frac{1}{2})$ and  $\mathbf{r}_2 = a(\frac{\sqrt{3}}{2}, \frac{1}{2})$ are the in-plane lattice vectors, $a$ is the lattice constant, and $\alpha,\beta$ are numbers between 0 and 1.  The stacking geometries induced by various interlayer translations $\mathbf{r}$ can be plotted on a real-space unit cell for each angular orientation (Figure \ref{StackGeom}). High symmetry stacking geometries are labelled by pairs or quadruplets of letters.  In our labelling convention, `A' refers to a metal atom and `B' to a chalcogen atom. A pair of letters in the stacking designation indicates sites aligned in the top and bottom layers, respectively.  For example ``AA" refers to a stacking geometry where metals in the top layer are aligned with metals in the bottom layer (A site over A site), and ``ABBA" refers to a geometry where a top layer metal is aligned with bottom layer chalcogens and top layer chalcogens are aligned over a bottom layer metal (A site over B site and B site over A site).  Stacking designations with four letters have two pairs of sites aligned in the top and bottom layers, while stacking designations with two letters have only one pair of aligned sites.  The stacking geometry that corresponds to no interlayer translation $\mathbf{r}$$(0,0)$ is AABB for the 0$^{\circ}$-oriented bilayer and ABBA for a 60$^{\circ}$-oriented bilayer.  These geometries appear on all four corners of their respective 0$^{\circ}$ or 60$^{\circ}$ stacking spaces (Figure \ref{StackGeom}). 

 \begin{figure*}
  \includegraphics[angle=0,width=\textwidth]{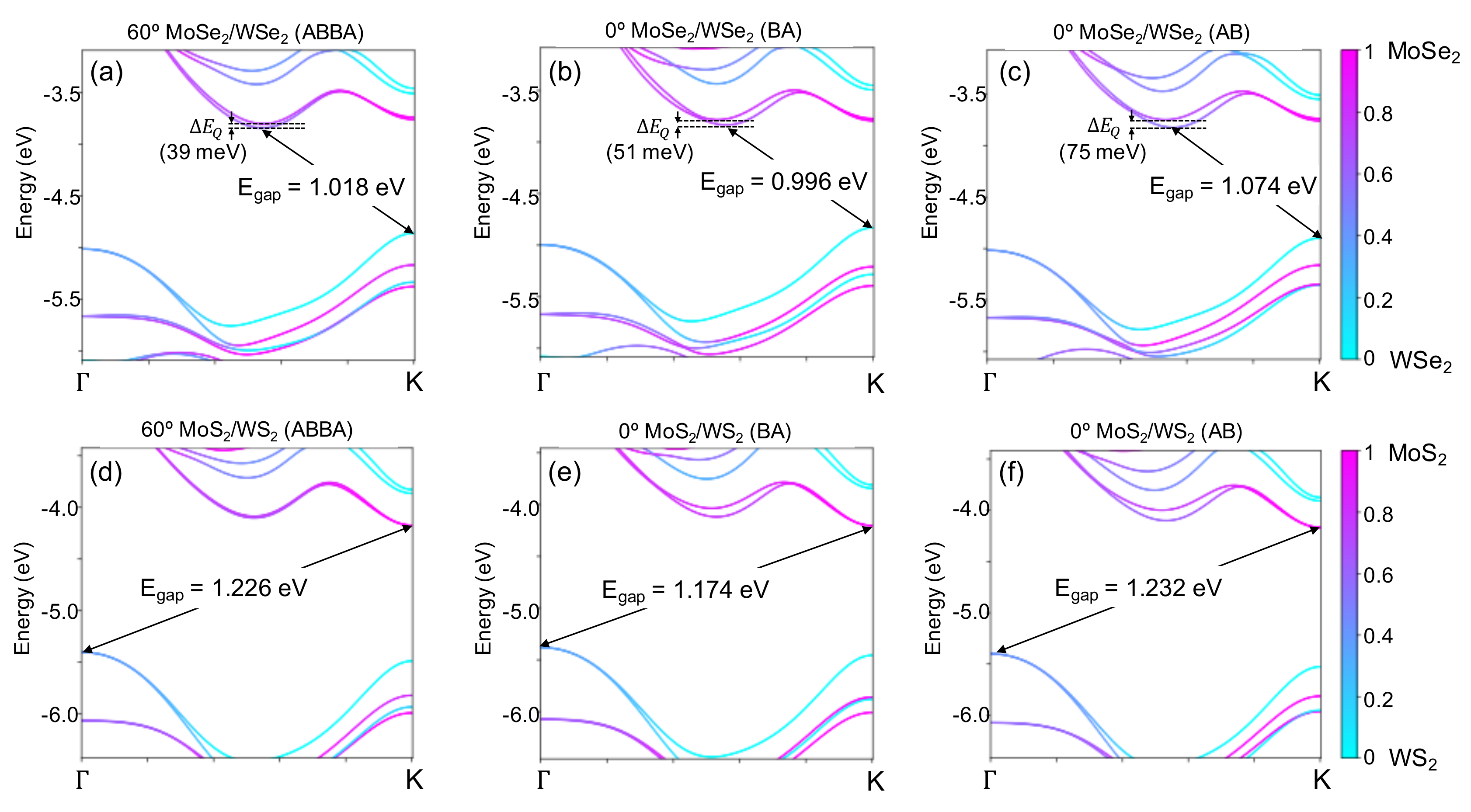}
  \caption{\label{lowEbands} Vacuum-aligned band structures of the ground state stacking geometries in (a) $60^{\circ}$-aligned MoSe$_2$/WSe$_2$ and (b), (c) $0^{\circ}$-aligned MoSe$_2$/WSe$_2$, as well as (d) $60^{\circ}$-aligned MoS$_2$/WS$_2$ and (e), (f) $0^{\circ}$-aligned MoS$_2$/WS$_2$.  Bands are plotted from $\Gamma$ to K.  The color scale indicates the layer weight of each state, with cyan corresponding to all weight coming from the WX$_2$ layer and magenta corresponding to all weight coming from the MoX$_2$ layer (X=S, Se). The selenide bilayers exhibit an indirect gap from K to Q, while the sulfide bilayers have an indirect gap from $\Gamma$ to K.  The degenerate ground state stacking geometries in the $0^\circ$-oriented systems (BA and AB) have distinct band gaps and conduction band Q-point splittings.  The vacuum-alignment of the band structures allows us to compare absolute band edge positions between different geometries within a given material.}
\end{figure*} 

We computed the vacuum-aligned band structures of the high symmetry stacking geometries in the $0^{\circ}$ and $60^{\circ}$-oriented MoS$_2$/WS$_2$ and MoSe$_2$/WSe$_2$ bilayers, since these are the extrema in stacking energy landscapes.  For the sulfide and selenide heterobilayers we study, the ground state stacking for the $60^\circ$-aligned bilayers is ABBA, and the $0^\circ$-aligned bilayers have two degenerate stacking geometries: AB and BA.  The highest energy stacking is AABB for the $0^\circ$-aligned bilayers and BB for the $60^\circ$-aligned bilayers.  In the $60^\circ$-aligned sulfide bilayer, AA stacking is a local minimum in energy\cite{Rosenberger2019}.

In the following, we focus on the ground state stacking geometries for two reasons. First, ground state stackings are expected to dominate the reconstructed moir\'e patterns of small angle twisted bilayers\cite{Carr2018}. Second, the minimum band gap in each configuration space occurs in a low-energy structure, and we expect excitons in moir\'e bilayers to drift towards the region of minimum gap.  Band structures for non-ground state high symmetry stackings are shown in the Supplemental Material\cite{Supp}.  All structures were relaxed with the metal atoms fixed in the x-y plane and allowed to relax in the z-direction (along the layer normal).  Chalcogens were allowed to relax in all three cartesian directions.  The lattice constants used in each system were $a$(S, $0^\circ$)=3.167 $\mathring{A}$, $a$(S, $60^\circ$)=3.166 $\mathring{A}$, $a$(Se, $0^\circ$)=3.295 $\mathring{A}$, and $a$(Se, $60^\circ$)=3.293 $\mathring{A}$, where each value was computed by minimizing energy with respect to lattice constant in the BA structure for $0^\circ$ degree oriented bilayers and in the ABBA structure for $60^\circ$ degree oriented bilayers. The difference between optimized lattice constants for the high symmetry geometries is less than 0.1\% in every case, so we judge that it is reasonable to use the same lattice constant for each geometry in a given stacking space. 

We computed band structures that include the effects of spin-orbit interactions. Inversion symmetry is broken in heterobilayers, and the presence of the heavy element tungsten makes the spin-orbit splittings significant. Because we use GGA functionals in our DFT calculations, the band gaps we report are significantly smaller than expected experimental values.  However, energy differences are expected to be more reliable, and we focus on these in the results that follow\cite{Kormanyos2015, Kang2013}.  
 
\begin{figure*}
  \includegraphics[angle=0,width=\textwidth]{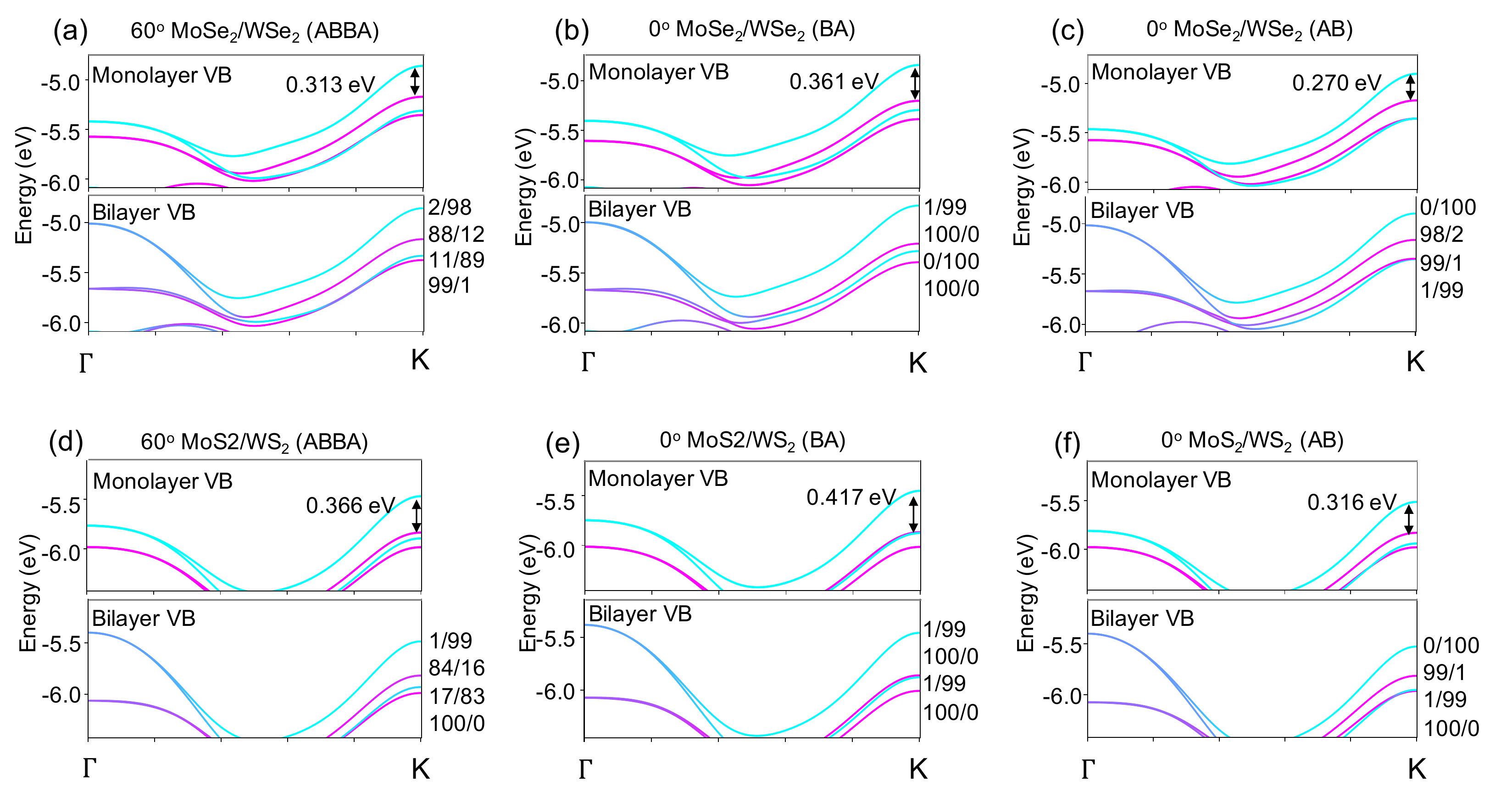}
  \caption{\label{BandsMLandHyb}  Valence bands for monolayers and bilayers in the low energy structures. The top half of each panel shows the monolayer valence bands, which have been aligned to the corresponding bilayer band structure. The bottom half of each panel shows the bilayer valence bands.  The color scale is the same as that in Figure \ref{lowEbands}, with the magenta bands corresponding to states localized on the MoX$_2$ layer and the cyan bands corresponding to states localized on the WX$_2$ layer.  The monolayer bands are aligned to the corresponding bilayer structure by matching the conduction band edge at K in MoX$_2$ to the conduction band edge at K in the bilayer and the conduction band edge at K in WX$_2$ to the second band above the conduction band edge in the bilayer, since the lowest four conduction bands in each bilayer at K are ``pure" states with no interlayer hybridization (See Figure \ref{lowEbands}.)  The bottom half of each panel shows the degree of valence band hybridization at K in each geometry.  The pairs of numbers X/Y indicate that for a given band, the corresponding state at K has X\% of its weight on the MoX$_2$ layer and Y\% of its weight in the WX$_2$ layer.}
\end{figure*}

 Figure \ref{lowEbands} shows vacuum-aligned band structures of the ground state stacking geometries of MoSe$_2$/WSe$_2$ and MoS$_2$/WS$_2$ bilayers, with the band gap size labeled for all band structures and the Q-point spin-orbit splitting labeled on the selenide band structures. All of the ground state structures have Type II band alignment at the K point, with the states at the valence band edge localized in the WX$_2$ layer and the states at the conduction band edge localized in the MoX$_2$ layer. The four lowest conduction bands at K are layer-polarized, i.e. there is no hybridization between layers.  In the valence band, the band edge corresponds to states well-localized in the WX$_2$ layer, but the second through fourth valence bands have different interlayer hybridizations depending on the structure. Most notable are the differences in hybridization between the ground state structures in the $0^\circ$ vs. $60^\circ$-oriented bilayers.  The third and fourth valence bands in the AB and BA structures are well layer-polarized, even though bands of the same spin are quite close in energy, whereas there is significant mixing between layers in the second and third valence bands in the ABBA structure (Figure \ref{BandsMLandHyb}).
  
\begin{figure*}
  \includegraphics[angle=0,width=\textwidth]{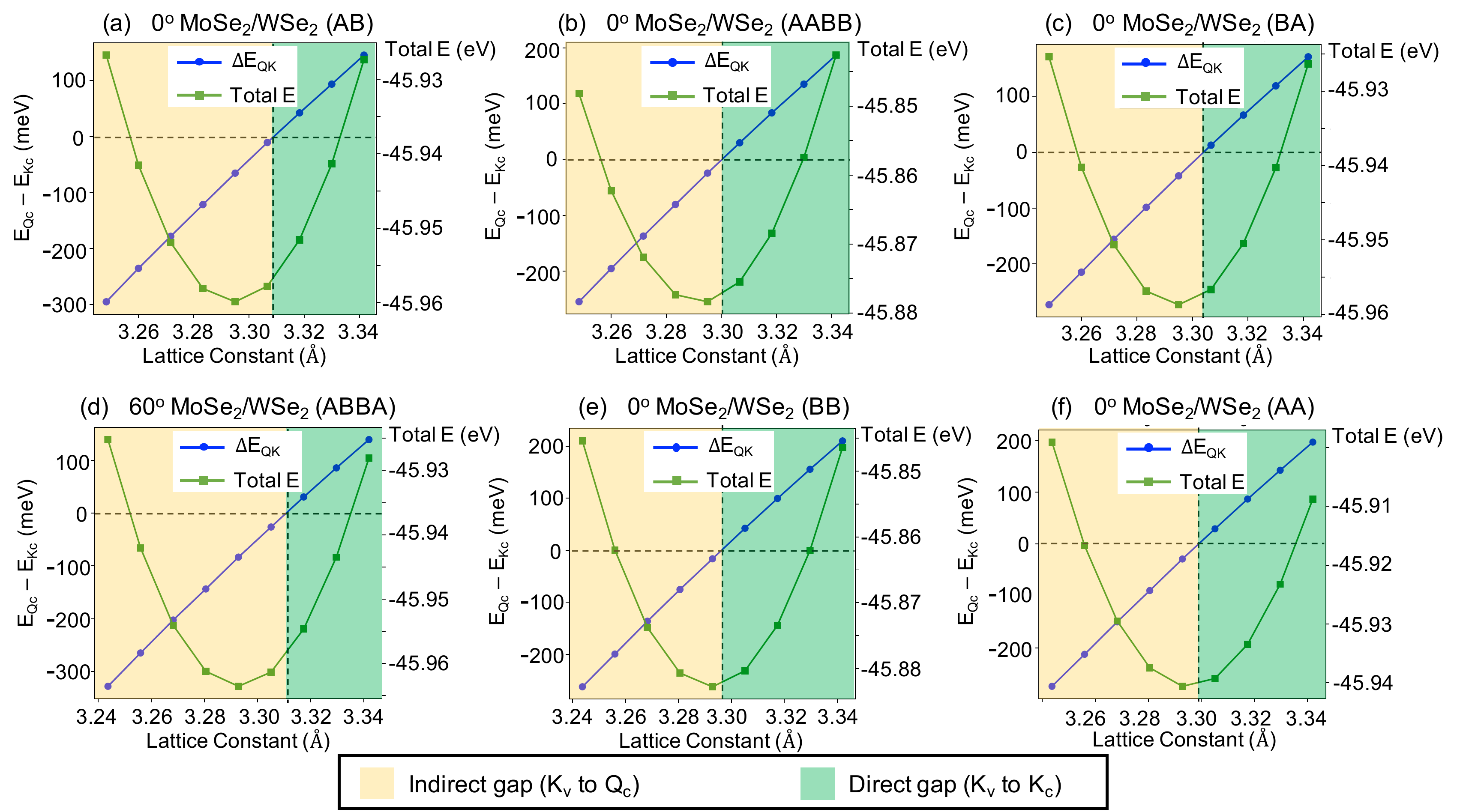}
  \caption{\label{LatticeConstVsGap} The indirect or direct nature of the band gap in the selenides depends sensitively on the choice of lattice constant. Here we plot the energy difference between conduction band minima at the K point and the Q point (blue curve) and the total energy (green curve) against lattice constant for the high symmetry stackings of $0^\circ$ and $60^\circ$-aligned MoSe$_2$/WSe$_2$ bilayers.  The optimized lattice constant for each structure is the value at the minimum of the total energy parabola.  The parabolic shape of the energy curve shows that we are straining the bilayers in the elastic regime. Lattice constants for which a structure has an indirect gap from K in the valence band to Q in the conduction band are highlighted in yellow, while lattice constants for which a structure has a direct gap at K are highlighted in green.  Each structure has an indirect gap at the optimized lattice constant, and the gap becomes direct for tensile biaxial strain greater than (a) $0.46\%$ (b) $0.24\%$ and (c) $0.30\%$ for the $0^\circ$ AB, AABB, and BA structures and (d) $0.55\%$ (e) $0.12\%$ and (f) $0.12\%$ for the $60^\circ$ ABBA, BB, and AA structures, respectively.}
\end{figure*}

 The minimum gap in the MoSe$_2$/WSe$_2$ ground states is indirect from the K to the Q point, where Q is a point lying between K and $\Gamma$.  While the indirect nature of the selenide bilayer band gap does depend on strain (Figure \ref{LatticeConstVsGap}), literature values for the thermal expansion coefficient\cite{SevikTEC} suggest that temperature effects are not enough to induce a direct gap at the sub-100K temperature scales in recent experiments \cite{TranMoire2019, SeylerMoire2019}.  The band gaps for the ground state structures are still squarely in the indirect gap regime at temperatures near 100K, while the AA and BB structures in the $60^\circ$-aligned bilayer are very close to being direct gap. In our 0K calculations, both the gap size and the magnitude of the spin-orbit splitting at the Q point differ among the three ground state structures, as shown in Figure \ref{lowEbands} (a)-(c).  Even the degenerate $0^{\circ}$ stackings AB and BA have band gaps that differ by 78 meV and Q point splittings that differ by 24 meV, which are energy scales accessible in optical experiments.

 In the ground state stacking geometries of the MoS$_2$/WS$_2$ bilayer, the band gap is indirect, with the valence band maximum at the $\Gamma$ point and the conduction band minimum at the K point.  As in the selenides, the band gap varies across geometries, but curiously, the $60^{\circ}$-aligned ground state (ABBA) and one of the $0^{\circ}$-aligned ground states (AB) have band gaps of the same size, while the other $0^{\circ}$-aligned ground state (BA) has a band gap 58 meV smaller than its partner.

\section{Discussion}

We study exclusively commensurate stackings of TMD heterobilayers in this work, so in some ways the structures we consider are all very similar.  However, the results reported above highlight a wide variety of both similarities and differences between the electronic structures of the various geometries.  For instance, we find that the two energetically degenerate ground states in the $0^\circ$ stacking space have very different band gaps.  This is true whether the chalcogens in the bilayer are sulfurs or seleniums, implying that there is an important generic difference between the AB and BA structures.  In the discussion that follows, we identify some chemical and symmetry-related factors that contribute to the similarities and differences between various structures and materials. 


To understand the differences in band structure across distinct stacking geometries, it's instructive to compare bilayer bands to the constituent monolayer bands.  In homobilayer calculations, the bilayer and monolayer band structures would be identical in the absence of interlayer coupling.  Indeed, for homobilayers the offset between bilayer bands corresponding to different layers is a direct proxy for the strength of interlayer coupling \cite{YaoInterlayer2017}.  One of the challenges of interpreting heterobilayer band structures is distinguishing bilayer band structure features that arise from interlayer coupling from those that arise from the ``hetero-" nature of the bilayer (i.e. the different metal atoms in the two layers).  In the top half of each panel in Figure \ref{BandsMLandHyb}, we show the valence bands of aligned monolayers, where we have aligned the MoX$_2$ and WX$_2$ bands by matching the monolayer conduction band edges to the pure state conduction bands at K in various bilayer geometries. (See Supplemental Material for full aligned monolayer band structures\cite{Supp}.) In this way, we can compare the band offsets that arise from the difference in metal atoms without any interlayer coupling effects.  The bottom half of each panel in Figure \ref{BandsMLandHyb} shows the valence bands of the fully interacting bilayer with the hybridizations at the K point labelled. We use these hybridizations as a proxy for interlayer coupling strength.  Some combination of interlayer coupling, the difference in metal atoms, and symmetry considerations contributes to variations in band structure across stacking geometries, and we comment on these factors below. 

The effect of the metal asymmetry on band gap size in our heterobilayers is most apparent when comparing the AB and BA structures in the MoSe$_2$/WSe$_2$ bilayer.  The band gap in both structures is from the K point in the valence band to the Q point in the conduction band, and the overall difference in band gap between the two structures is 78 meV.  The difference in Q point minimum in AB vs. BA is only 9 meV, so most of the band gap difference arises from the difference in the valence band maxima (VBM) at K.  The difference in VBM cannot be a function of differences in interlayer metal-to-metal distance since this is roughly the same in the AB and BA structures (see Figure S6 \cite{Supp}), and it can't be a consequence of hybridization pushing bands apart since there is little to no hybridization at the top of the valence band at K in these structures (bottom sections of Figure \ref{BandsMLandHyb} (b), (c)).  We instead conjecture that the difference in band edge alignment between monolayer bands in AB as compared to BA is a consequence of the different dielectric environments of the metal atoms, since metal atom orbitals dominate the monolayer band edges at K \cite{Jones2014}.  In the AB structure, the W atoms are surrounded by a greater number of chalcogens, whereas in the BA structure, the Mo atoms are surrounded by the greater number of chalcogens.  We can estimate the size of this effect by measuring the monolayer band offsets at the VBM in the aligned monolayers in Figure \ref{BandsMLandHyb}.  The difference in monolayer band offsets is about 91 meV between the AB and BA structures, with the band offset higher in BA, which has the smaller bilayer band gap.  The argument that these band offsets come from the metal atom asymmetry in our bilayers is strengthened by the fact that the difference between monolayer band offsets at K in the AB and BA structures of the MoS$_2$/WS$_2$ is roughly the same (within 10 meV) as the selenide bilayer, despite the difference in chalcogen.  

Interlayer hybridization can also explain some of the variation in band structures across geometries.  For instance, in the selenide bilayers it is the hybridization of bands at the Q point that pushes the Q-point conduction bands below the bands at K, making the selenide heterobilayers indirect gap materials in contrast to the direct gap monolayer MoSe$_2$.  In the sulfide bilayers, it is the hybridization of the valence bands at $\Gamma$ that gives the bilayers their indirect gap, as seen by comparing the aligned monolayer and bilayer valence bands in Figure \ref{BandsMLandHyb}. (The exception is the high energy BB and AABB sulfide structures, where hybridization at $\Gamma$ is not strong enough to shift the $\Gamma$ point higher than the K point, probably due to the large interlayer distance in the high energy stackings. See Figures S1 and S2\cite{Supp}.) 

Interlayer hybridization and the splitting of bands that results is a function of interlayer distance and the energy offset of the unhybridized bands \cite{XuHetReview2018}. However, the alignment of a system can also effect the mixing, a fact exemplified by the differences in hybridization near the valence band edge at K in ABBA vs. AB and BA structures.  The orbitals at the valence band edges are predominantly metal d orbitals, and for the specific metal atom symmetry of the ABBA structure, there should be finite mixing between Mo and W bands at the valence band edge at K \cite{Jones2014}.  Indeed we find finite hybridization between the valence bands with the same spin at K in both the sulfide and selenide bilayers (Figure \ref{BandsMLandHyb} (a), (d) ).  The AB and BA structures have the same metal atom symmetry as ABBA, so we would naively expect to find finite mixing in the valence band at K in both structures. However sulfide and selenide AB and BA heterobilayers all show negligible interlayer mixing between valence bands with the same spin at K (Figure \ref{BandsMLandHyb} (b)-(c), (e)-(f) ).  This discrepancy arises from the distinct Brillouin zone alignments in $0^\circ$ vs. $60^\circ$ structures. In the $0^\circ$-aligned AB and BA structures, the K points are aligned, while in the $60^\circ$-aligned ABBA structure, the K point in one layer is aligned with the K' point in the other layer.  Since the K and K' points are time-reversed partners, the wavefunctions at these points are related by complex conjugation. This difference in the phase of the wavefunctions at K and K' causes the contribution to the interlayer hopping term to cancel out in the $0^\circ$-aligned structure, forbidding hybridization at the level of the metal-atom tight-binding model.

The differences in band gap size and location as well as spin-orbit splitting in the various stacking geometries should have consequences for photoluminescence (PL) of the heterobilayers.  Hanbicki, et al. argue \cite{HanbickiILE} that the splitting of the interlayer exciton (ILE) peak in PL measurements corresponds to the spin-orbit splitting at the Q point of the bilayer electronic structure.  If this is the case, then different geometries could be distinguished by ILE peak splitting as well as by ILE peak location.  For example, the PL signal for a MoSe$_2$/WSe$_2$ bilayer in the AB stacking geometry would have an ILE peak at an energy about 78 meV higher than a bilayer in the BA stacking geometry, and the peak splitting would be 24 meV larger for the AB stacked bilayer.  Furthermore, the strain-dependence of the band gaps could have an effect on optical experiments, particularly in moir\'e bilayers, where strain effects can be significant.  Under sufficient tensile strain, selenide heterobilayers will become direct gap, which might quench the ILE signal altogether, since the wavefunctions at the valence band maximum and conduction band minimum at K are in separate layers and thus have no overlap.   

\section{Conclusion}

One route to understanding experiments on twisted TMD heterostructures is to understand the properties of the constituent commensurate regions of such moir\'e structures.  In this paper, we used first-principles calculations to compute electronic structure of commensurately stacked MoSe$_2$/WSe$_2$ and MoS$_2$/WS$_2$ bilayers in both the $0^\circ$ and $60^\circ$ orientations.  We identify variations in band gap and spin-orbit splitting, which arise from the metal atom asymmetry of the heterobilayers and differences in interlayer hybridization.  These results may allow the stacking structure of heterobilayers to be probed using optical techniques. 

\section*{Acknowledgements}
We acknowledge H.-J. Chuang, A. T. Hanbicki, B. T. Jonker, I. I. Mazin, K. M. McCreary, M. R. Rosenberger, S. V. Sivaram,  and D. Wickramaratne for helpful discussions.  This research was performed while M.P. held a National Research Council fellowship at NRL.  This work was supported by core programs at NRL and the NRL Nanoscience Institute.  Computational work was supported by a grant of computer time from the DoD High Performance Computing Modernization Program at the U.S. Army Research Laboratory and the U.S. Air Force Research Laboratory Supercomputing Resource Centers.

\bibliography{TMDvdwPaper2}

\begin{thebibliography}{10}
\bibitem{Notation}
Giving a monolayer the ``H" label is something of an abuse of notation, since H
  usually refers to the $60^\circ$ stacking between multiple monolayers, as in
  bilayer 2H TMDs. Here we use the H designation to indicate that the relative
  positions of metal and chalcogens within a monolayer is the same as that
  found in the monolayer components of 2H TMDs. This distinguishes our
  monolayers from 1T or 1T' monolayers, where the chalcogen atoms are displaced
  from the positions they occupy with respect to the metal atoms in 2H phase
  (or indeed 3R and bulk) TMDs.
\end{thebibliography}
\bibliographystyle{ieeetr}
\end{document}


\title{Supplement to Commensurate structures in twisted transition metal dichalcogenide heterobilayers}
\author{Madeleine Phillips and C. Stephen Hellberg}
        \affiliation{Naval Research Laboratory, Washington, D.C. 20375\\}

\maketitle

\begin{figure}
  \includegraphics[angle=0,width=0.95\columnwidth]{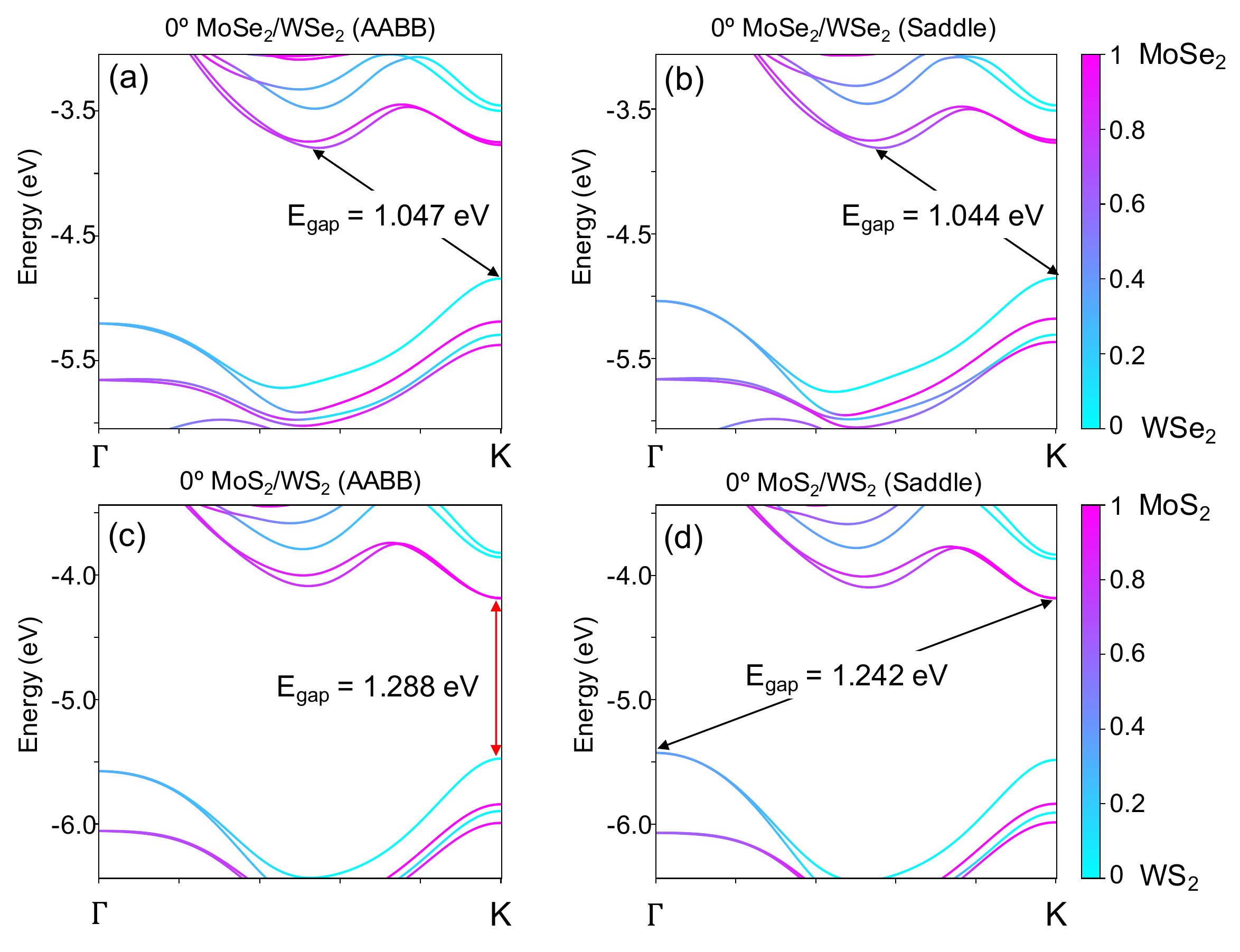}
  \caption{\label{0BandsSupp} Remaining high symmetry band structures for $0^\circ$-aligned (a)-(b) MoSe$_2$/WSe$_2$ and (c)-(d) MoS$_2$/WS$_2$ bilayers.  AABB is the high energy structure at $(\alpha, \beta) = (0, 0)$ and the Saddle point structure is halfway between the AB and BA structures at $(\alpha, \beta) = (1/2, 1/2)$.  The band structures in both materials have a type II band alignment at K.  As in the other selenide bilayer structures, the AABB and Saddle point stackings have an indirect gap from K in the valence band to Q in the conduction band. The saddle point structure in the sulfide bilayer has an indirect gap from $\Gamma$ in the valence band to K in the conduction band, like the other sulfide band structures.  However, the high energy AABB sulfide bilayer has a direct gap at K, with a magnitude larger than any other $0^\circ$-aligned sulfide bilayer band gap in this study.  }
\end{figure}

\begin{figure}
  \includegraphics[angle=0,width=0.95\columnwidth]{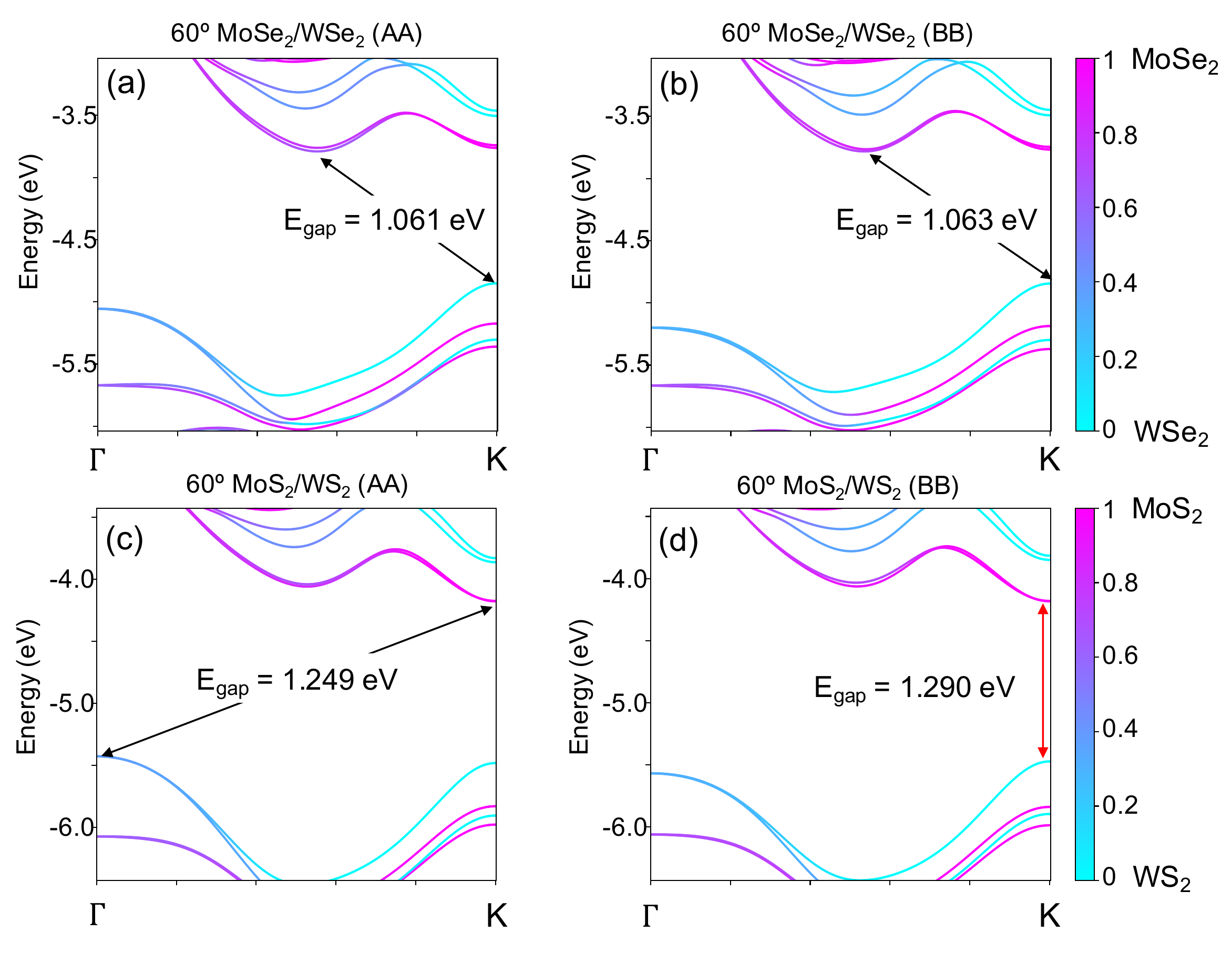}
  \caption{\label{60BandsSupp} Remaining high symmetry band structures for $60^\circ$-aligned (a)-(b) MoSe$_2$/WSe$_2$ and (c)-(d) MoS$_2$/WS$_2$ bilayers.  AA is a structure located at $(\alpha, \beta) = (2/3, 1/3)$ in the $60^\circ$ configuration space.  In the sulfide bilayer, AA is a local minimum, and indeed some calculations of sulfide homobilayers like MoS$_2$/MoS$_2$ find the AA structure to be the global minimum \cite{Carr2018}. The BB structure is the high energy structure in the $60^\circ$ configuration space, located at $(\alpha, \beta) = (1/3, 2/3)$.  The band structures in both materials have a type II band alignment at K.  As in the other selenide bilayer structures, the AA and BB stackings have an indirect gap from K in the valence band to Q in the conduction band. The AA structure in the sulfide bilayer has an indirect gap from $\Gamma$ in the valence band to K in the conduction band, like the other sulfide band structures.  However, the high energy BB sulfide bilayer has a direct gap at K, with a magnitude larger than any other $60^\circ$-aligned sulfide bilayer band gap in this study. }
\end{figure}

\begin{figure}
\includegraphics[angle=0,width=0.95\columnwidth]{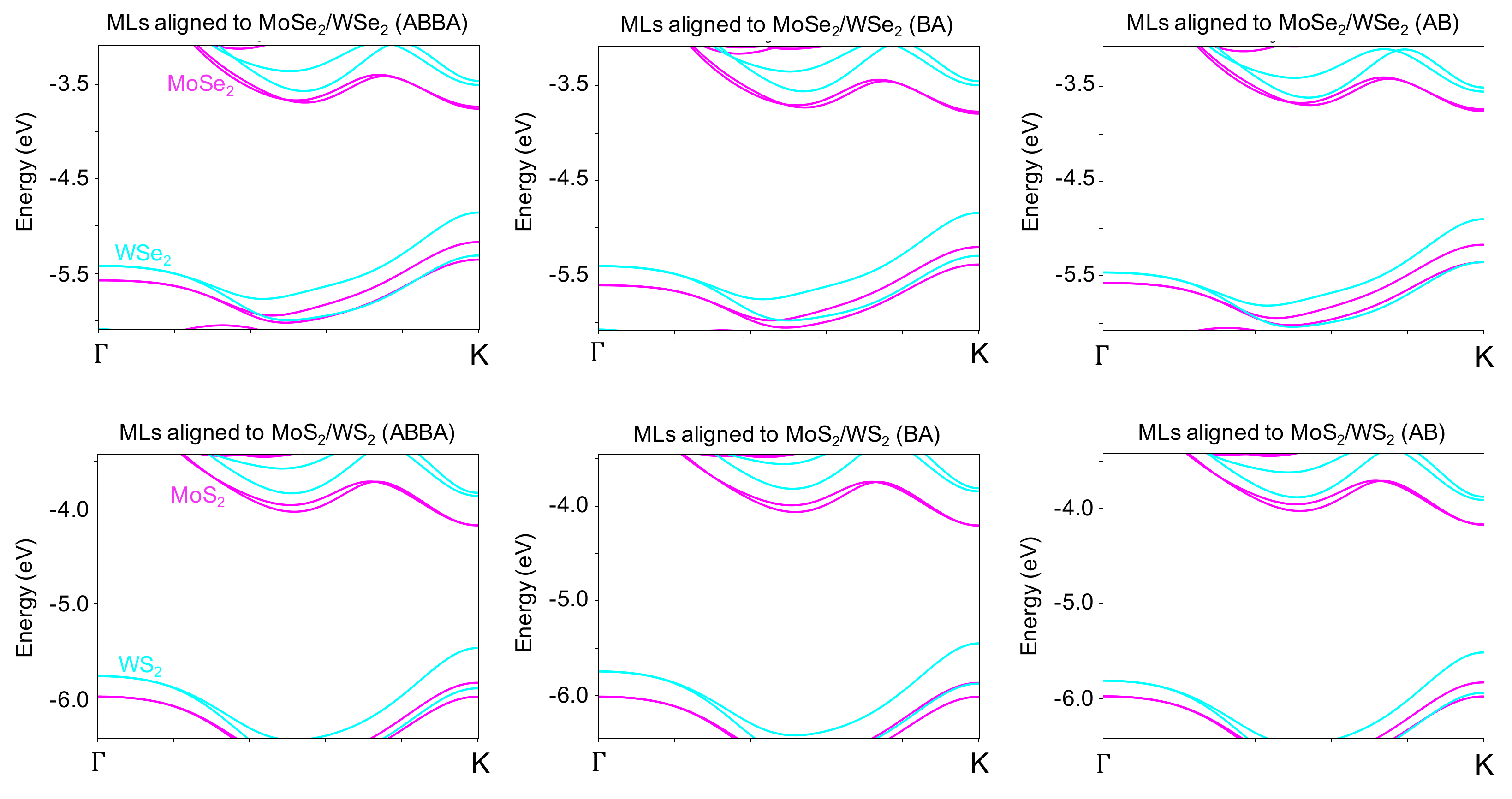}
  \caption{\label{MLsLowE}   Aligned monolayer bands for ground state structures in $0^\circ$ and $60^\circ$ stacking spaces.  The conduction band edge of each monolayer at K is aligned with the appropriate bilayer pure state at K. (i.e. The MoX$_2$ conduction band edge is aligned with the bilayer conduction band minimum, which is entirely localized in the MoX$_2$ layer, and the WX$_2$ conduction band edge is aligned with the third lowest bilayer conduction band at K, which is entirely localized on the WX$_2$ layer.) Magenta bands are the MoX$_2$ bands and cyan bands are the WX$_2$ bands. }
\end{figure}

\begin{figure}
\includegraphics[angle=0,width=0.95\columnwidth]{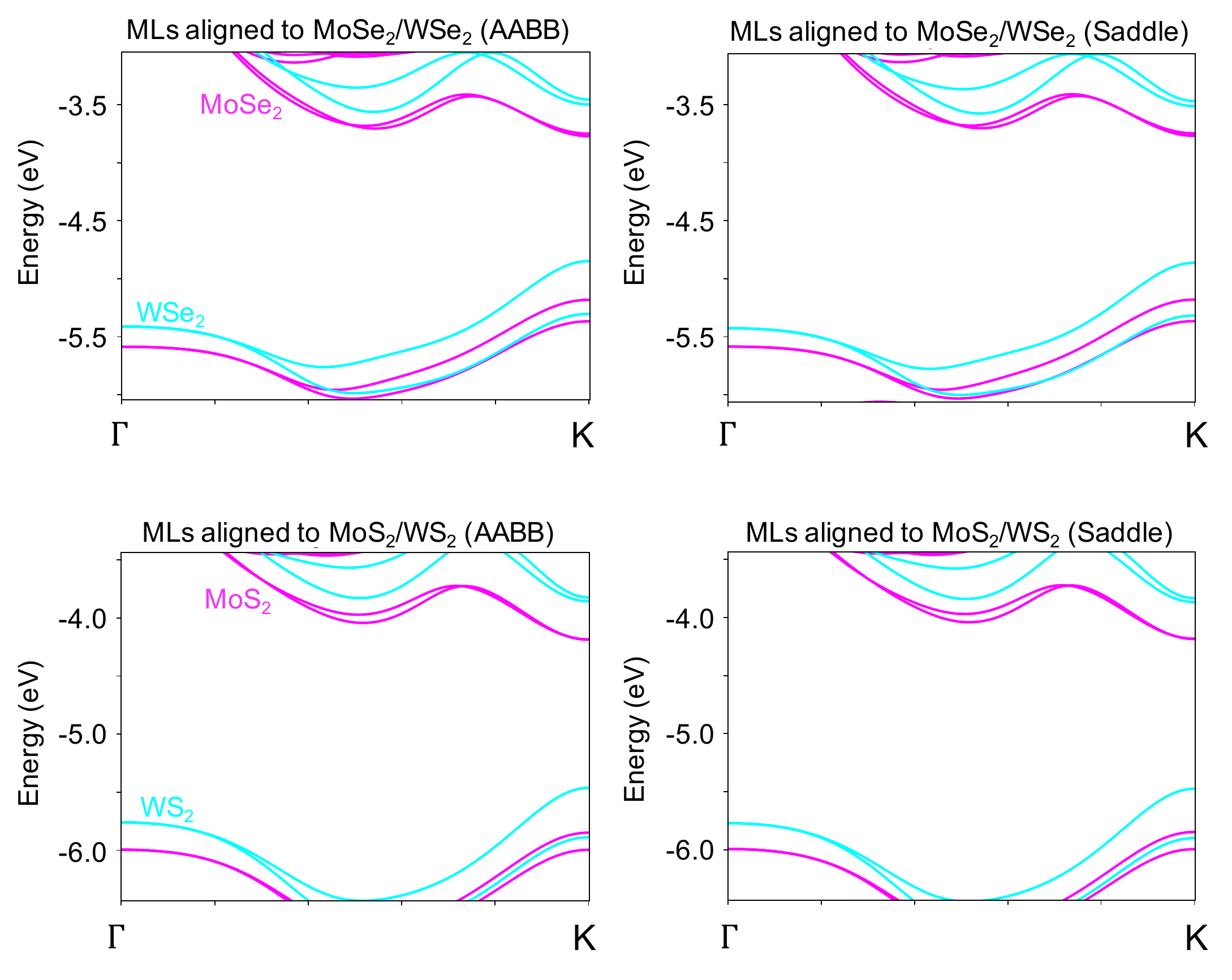}
  \caption{\label{MLs0_high}  Aligned monolayer bands for the higher energy structures in the $0^\circ$ stacking space.  The conduction band edge of each monolayer at K is aligned with the appropriate bilayer pure state at K. (i.e. The MoX$_2$ conduction band edge is aligned with the bilayer conduction band minimum, which is entirely localized in the MoX$_2$ layer, and the WX$_2$ conduction band edge is aligned with the third lowest bilayer conduction band at K, which is entirely localized on the WX$_2$ layer.) Magenta bands are the MoX$_2$ bands and cyan bands are the WX$_2$ bands.}
\end{figure}

\begin{figure}
\includegraphics[angle=0,width=0.95\columnwidth]{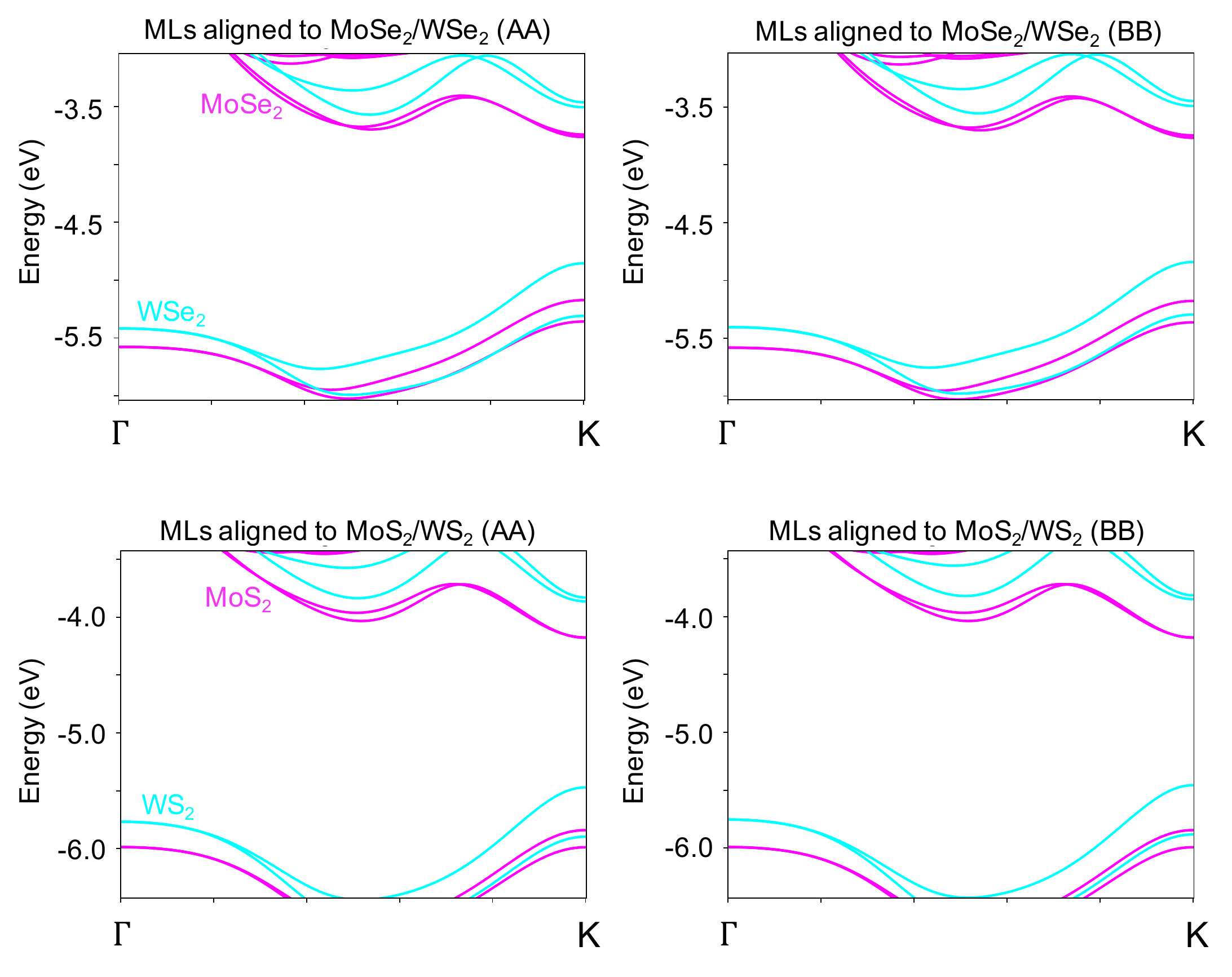}
  \caption{\label{MLs60_high}  Aligned monolayer bands for the higher energy structures in the $60^\circ$ stacking space.  The conduction band edge of each monolayer at K is aligned with the appropriate bilayer pure state at K. (i.e. The MoX$_2$ conduction band edge is aligned with the bilayer conduction band minimum, which is entirely localized in the MoX$_2$ layer, and the WX$_2$ conduction band edge is aligned with the third lowest bilayer conduction band at K, which is entirely localized on the WX$_2$ layer.) Magenta bands are the MoX$_2$ bands and cyan bands are the WX$_2$ bands.}
\end{figure}

\begin{figure*}
  \includegraphics[angle=0,width=\textwidth]{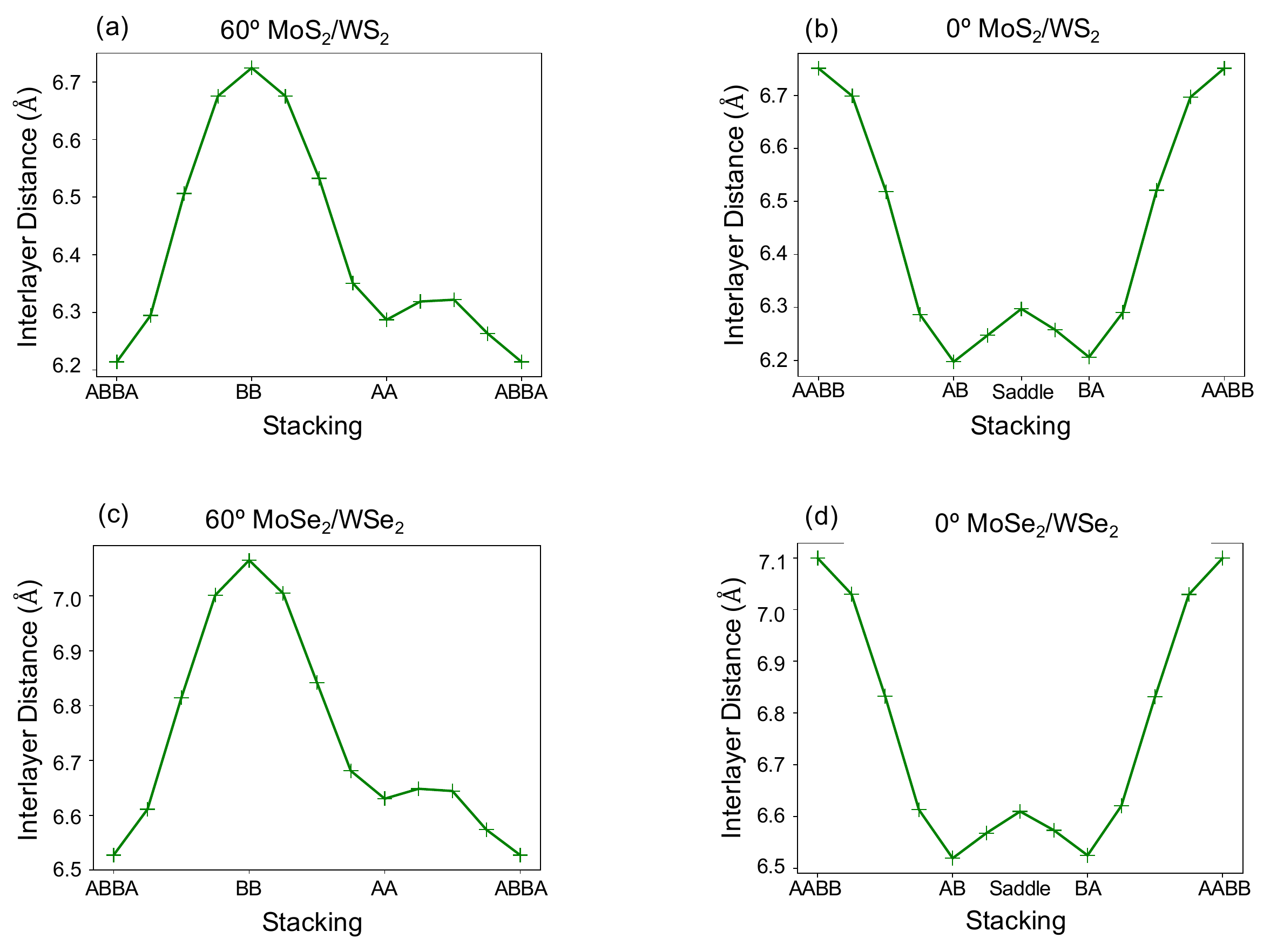}
  \caption{\label{ILD} Interlayer distances plotted with respect to stacking geometries for (a) $60^{\circ}$ and (b) $0^{\circ}$-aligned MoS$_2$/WS$_2$ bilayers and (c) $60^{\circ}$ and (d) $0^{\circ}$-aligned MoSe$_2$/WSe$_2$ bilayers.  Interlayer distance is measured between the metal atoms of the two layers. The difference in interlayer distance between the degenerate ground states (AB and BA) in $0^\circ$-oriented MoS$_2$/WS$_2$ is $0.009 \mathring{A}$, while the difference in interlayer distance between the degenerate ground states (AB and BA) in $0^\circ$-oriented MoSe$_2$/WSe$_2$ is $0.006 \mathring{A}$.}
\end{figure*}

\bibliography{TMDvdwPaper2}
\bibliographystyle{ieeetr}

%
%